\documentclass[prl,aps,showpacs,floatfix]{revtex4}
\usepackage{amsmath}
\usepackage{graphicx}
\begin{document}
\title{Negativity of the Coarse Grained Wigner Function as a Measure of 
Quantal Behavior}
\author{Tyler E. Keating, Adam T.C. Steege, and Arjendu K. Pattanayak}
\date{\today}
\affiliation{Department of Physics and Astronomy, Carleton College, \\ 1 North
College Street, Northfield, MN 55057}
\begin{abstract}
The negativity of a given state's Wigner function has been proposed
as a measure of quantumness of that state in a unipartite system. 
This otherwise physically intuitive and useful phase-space measure however 
does not yield the right correspondence principle limit, and also turns
out to yield infinite values of the infinite square well. We show that
both these issues can be sensibly resolved using coarse-graining of the
Wigner function.
\end{abstract}
\pacs{03.65.Ta}
\maketitle
\section{Introduction}
A numerical measure of the ``quantumness" of a system is useful in many
contexts, for example in nonlinear optics or studies of the transition 
from quantum-to-classical behavior. There are various ways of
parameterizing nonclassicality. It is typical to start by defining a 
quantum coherent state with minimum uncertainty as the most classical system 
possible, then measuring how different the system in question is from such 
a coherent state. The difference between these two states is often quantified 
by finding the minimum distance between them based on the trace, the 
Hilbert-Schmidt distance, or some other similar metric\cite{dodonov}.  
Alternatively, the distance between a given system and the closest 
classical system can be measured using the Cahill parameter $\tau$. The 
minimum value of $\tau$ which produces a positive definite distribution 
function can be interpreted as the "nonclassical depth" of the system, 
which in turn may be used as an index of quantumness\cite{lee}.
A more thorough discussion of many of the nonclassicality parameters
that have been studied can be found in Dodonov\cite{dodonov}.

Recently, it has been argued\cite{kenfack} that the negativity of a 
given state's Wigner function is a simple and clean measure of that state's
quantumness. One advantage of this method is that the Wigner
representation can be determined experimentally\cite{royer}, as well as 
the fact that the negativity is a measure of non-locality for bi-partite 
or multi-partite systems. Further, for a unipartite system, this is
an intuitive measure. Since Wigner functions are quasi-probabilities 
existing in phase-space, they are directly comparable with classical 
probability functions, and this is often done in understanding 
quantum-classical correspondence issues, for example.  There is no 
negativity in classical probablities, of course.  The negativity, or 
negative volume, of a Wigner function $W(x,p)$ defined as:
\begin{equation}
\eta=\int_{-\infty}^{\infty}\int_{-\infty}^{\infty}\frac{\left|W(x,p)\right
|-W(x,p)}{2}dxdp
\end{equation}
Since the parts of the Wigner function that are negative decrease as the 
system decoheres under the influence of an environment, this further 
reinforces the notion that the Wigner function's negativity is a useful
measure.

In the case of the one-dimensional harmonic oscillator, $\eta$ increases 
approximately as $\sqrt{n}$ when measured for eigenstates of different 
$n$\cite{kenfack}. At low values of $n$, this is sensible -- we expect 
that the increased oscillations as we climb the eigenfunction ladder do 
indeed correspond to greater quantumness. However, this does not hold 
at high values of $n$. The correspondence principle dictates that the limit 
as $n\rightarrow\infty$ of the harmonic oscillator\cite{segatto} should 
yield a transition from quantum to classical behavior. As such, a monotonic 
increase in quantumness with $n$ is not correct. Moreover, simple intuition 
runs counter to the possibility that \emph{any} system becomes extremely 
quantum at the high energies associated with the macroscopic world; even 
if the quantumness of such a system did not drop off to zero at high $n$, 
its higher energy states should at the very least not be orders 
of magnitude \emph{more} quantum than the lower states where quantum 
behavior is usually observed.

As a way of dealing with this counter-intuitive aspect of the otherwise
sensible measure, we propose `coarse-graining' the Wigner function by 
convolving it with a Gaussian before measuring its negativity as
\begin{align}\label{coarse}
\eta&=\int_{-\infty}^{\infty}\int_{-\infty}^{\infty}\frac{\left|W_{CG}(x,p)
\right|-W_{CG}(x,p)}{2}dxdp,
\\
\text{where }
W_{CG}(x,p)&=\int_{-\infty}^{\infty}\int_{-\infty}^{\infty}
\exp[-\delta((x-x')^2+(p-p')^2)]W(x' ,p')dx'dp'.
\end{align}
Mathematically, convolution with a Gaussian has the effect of smoothing 
out the Wigner function, reducing the magnitude of local oscillations.
This physically motivated technique that has often been used to study 
semiclassical behavior in quantum systems. Coarse-graining has the
effect of smoothing away small and purely quantum features while maintaining 
the large-scale structure from which classical behavior typically 
emerges\cite{rivas}. 
As discussion in the literature shows\cite{habib,wilkie}, coarse-graining 
is necessary to retrieve classical behavior from some quantum systems.
It represents the reality that quantum-classical correspondence for closed 
quantum systems provides a singular classical limit -- by this we mean
that $\hbar =0$ can be qualitatively, not just quantitatively, different from 
any non-zero value of $\hbar$. If the negativity of the Wigner function is 
to be a meaningful measure of the classicality of such systems, it is
an intuitive step to include coarse-graining. In fact, as we show below, 
this procedure indeed resolves the paradoxical behavior(s) of the
negativity.

Moreover, as we show below, the infinite square well, another textbook
example, also benefits from this procedure. Without coarse-graining,
eigenfunctions of this system have infinite negativity, which again
renders this otherwise useful measure less useful, but coarse-graining
renders the negativity finite.

In what follows, we present our results for the harmonic oscillator,
followed by the infinite square well, and conclude with a short
discussion.

\section{The Harmonic Oscillator}
In arbitrary units where $m=\omega=\hbar=1$, the Wigner function for the 
$n^{th}$ eigenstate of a quantum harmonic oscillator is\cite{kenfack}
\begin{equation}
W_n(x,p)=2\frac{(-1)^n}{\pi}\exp[-2(x^2+p^2)]L_n[4(x^2+p^2)]
\end{equation}
where $L_n$ denotes the $n^{th}$ Laguerre polynomial. After coarse-graining 
this function as in Eq.~(\ref{coarse}), if $\delta$ is an integer, the 
resulting function takes the form 
\begin{equation}
W_{CG,n}(x,p)=ke^{\frac{2n}{\delta+2}(x^2+p^2)}P_n(x,p)
\end{equation}
where $k$ is an arbitrary numerical constant, and $P_n(x,p)$ is an $n^{th}$ 
order polynomial function of $(x^2+p^2)$. For example we have that 
\begin{align}
W_{\delta=3,n=4}=6.84\times 10^{-2}e^{-\frac{6}{5}(x^2+p^2)}[(x^2+p^2)^4-
\frac{20}{9} (x^2+p^2)^3+\frac{25}{18}(x^2+p^2)^2-\frac{125}{486}(x^2+p^2)+
\frac{625}{69984}]\notag
\\
W_{\delta=5,n=2}=0.947\times e^{-\frac{10}{7}(x^2+p^2)}[(x^2+p^2)^2-
\frac{21}{25}(x^2+p^2)+\frac{441}{5000}]\notag
\end{align}
For the special case $\delta=2$, which corresponds to 
$\Delta x\Delta p=\frac{\hbar}{2}$, this reduces to
\begin{equation}
W_{CG,2}(x,p)=\frac{1}{2n!}e^{-(x^2+p^2)}(x^2+p^2)^n
\end{equation}
which looks like a classical orbit. This function is positive definite, 
meaning that, as expected\cite{soto}, no quantumness can be 
observed with blurring greater than or equal to minimum uncertainty.
The Husimi function, a positive-definite quantum `probability' function
in phase-space is, in fact, constructed by coarse-graining the Wigner 
function with a minimum-uncertainty Gaussian. This result implies
that in this specific case the `least' amount of coarse-graining 
needed to get a function indistinguishable from classical is precisely 
the Husimi coarse-graining. Since the harmonic oscillator basis can be
used to construct an arbitrary state, this means that a general state
becomes positive definite if and only if a Husimi coarse-graining is
used.
\begin{figure}[htp]
\centering
\includegraphics{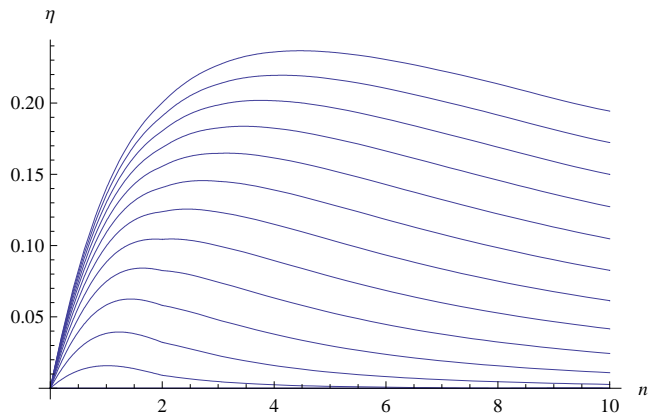}
\caption{Negativity $\eta$ of the harmonic oscillator's smoothed Wigner 
function at various $n$ shown for $\delta$'s increasing from 3 on the bottom 
to 14 on top. As $\delta$ increases, the peak in negativity moves to 
higher $n$.}
\label{curves}
\end{figure}
\begin{figure}[htp]
\centering
\includegraphics{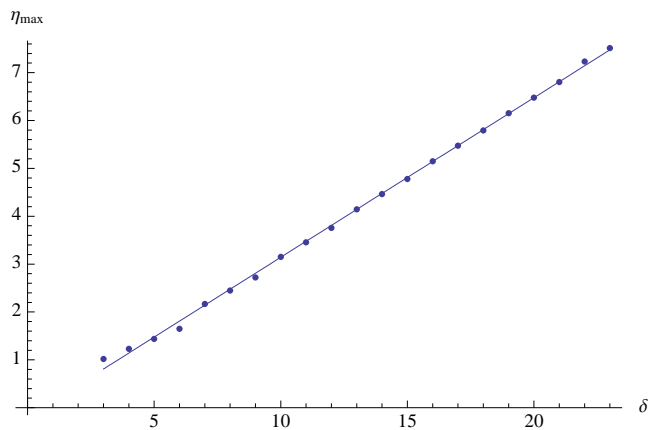}
\caption{The eigenfunction with maximum negativity $n_{max}$ increases 
monotonically and roughly linearly as a function of $\delta$.}
\label{Nmax}
\end{figure}

More interestingly, once the Wigner function is coarse-grained, 
its negativity no longer increases monotonically with $n$, and regardless 
of $\delta$, negativity does not \emph{decrease} monotonically with $n$ 
either. Rather, negativity increases sharply to some $n_{max}$, then tapers 
off to approach zero as $n\rightarrow\infty$ (Fig. 1). As $\delta$ increases, 
so does the $n$ value at which maximum negativity is observed. All of these 
properties are satisfyingly in line with physical intuition and provide a 
{\em post facto} justification for the coarse-graining procedure.
The relationship between $\delta$ and $n_{max}$ seems to be roughly linear, as 
can be seen in Fig.~(\ref{Nmax}). More specifically, the relation between
$\eta_{max}$ and $\delta$ is best fit by a line with slope $\approx 
\frac{1}{3}$ for which we have no qualitative explanation.
\begin{figure}[htp]
\centering
\includegraphics{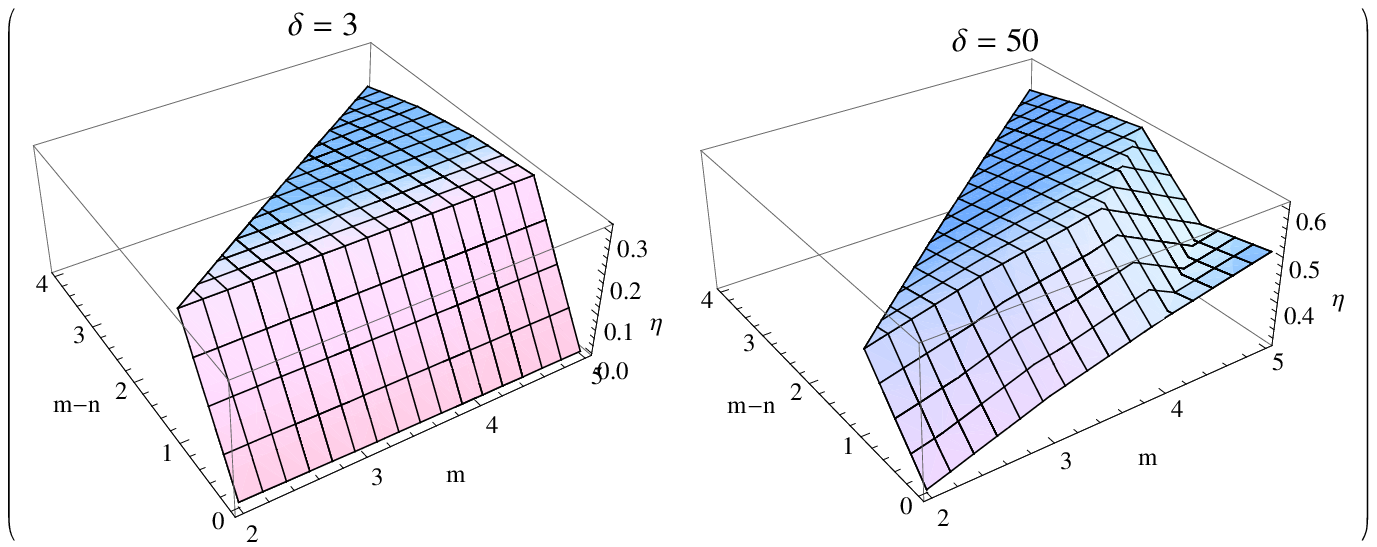}
\caption{Negativity $\eta$ of the harmonic oscillator's smoothed Wigner 
function for the state  $\left | m \right \rangle \left \langle n \right |$ 
shown as a function of $m$ and $m-n$ in both the $\delta=3$ and $\delta=50$ 
cases.}
\label{offdiags}
\end{figure}

In addition to the Wigner function for the $n^{th}$ eigenfunction, an
arbitrary state for the harmonic oscillator will have a Wigner function
with ``off-diagonal'' elements arising from the combination of two or more 
pure states. For instance, for a harmonic oscillator state 
$\left |\psi \right \rangle=\left | m \right \rangle+\left | n \right \rangle$, 
the overall wave function $\left | \psi \right \rangle \left \langle \psi 
\right |$ could be written as $\left | m \right \rangle \left \langle m \right 
|+\left | n \right \rangle \left \langle n \right |+\left | m \right \rangle 
\left \langle n \right |+\left | n \right \rangle \left \langle m \right |$, 
with $\left | m \right \rangle \left \langle n \right |$ and $\left | n \right 
\rangle \left \langle m \right |$ being the off-diagonal elements. In the 
same arbitrary units as above where $m=\omega=\hbar=1$, the Wigner functions 
of these off-diagonal elements are given by\cite{garraway}
\begin{align}
\label{offwig}
W_{m,n}(x,p)=&2\frac{(-1)^n}{\pi}\sqrt{\frac{n!}{m!}}\exp[-2(x^2+p^2)](2
\sqrt{x^2+p^2})^{m-n}*\notag
\\
&\cos((m-n)\arctan(\frac{p}{x}))L_n^{m-n}[4(x^2+p^2)]
\end{align}

Some of the negativities for various combinations of $n$ and $m$ at a few 
different values of $\delta$ are shown in Fig. 3. The complexity of equation 
\ref{offwig} makes a thorough analysis of the behavior of smoothed, 
off-diagonal Wigner functions computationally difficult. Nevertheless, a few 
features of Fig. 3 are worth pointing out. All values of $m-n$ show the same 
basic pattern of increasing negativity with increasing $m$, at least up to a 
point. It is also interesting that the $m-n=0$ cases (i.e. the on-diagonal 
elements) tend to have significantly lower negativities than any other values 
of $m-n$. This is commensurate with intuition that a superposition of
quantum states is in some sense a more non-classical state.
Irrespective the existence of any patterns, it is important to note that 
the off-diagonal Wigner functions can in principle be analytically 
coarse-grained. Since the harmonic oscillator states form a basis for 
all infinite states, any infinite Wigner function can then be expressed 
as a superposition of on- and off-diagonal square well states, and
therefore sensibly coarse-grained analytically before measuring its 
negativity.

\section{The Square Well}
The infinite square well provides another case in which coarse-graining 
before calculating negativity is a useful tool.  The Wigner function for the
$n$th eigenstate of an infinite square well of width $\pi$ is\cite{belloni}
\begin{equation}
\label{squarewig}
W(x,p)=\frac{2}{\pi^{2}\hbar}(\frac{\sin[2(p/\hbar+n)x]}{4(p/\hbar+n)}+ 
\frac{\sin[2(p/\hbar-n)x]}{4(p/\hbar-n)}-\cos(2nx)\frac{\sin[2px/\hbar]}
{2p/\hbar})
\end{equation}
where $x\in[0,\pi/2]$; for $x\in[\pi/2,\pi]$, the expression is identical, 
except all occurences of $x$ are replaced by $L-x$.  A particle in the 
infinite well can only occupy a finite range of positions but can have any 
momentum, so the support of this Wigner function 
covers $x\in[0,\pi]$ and $p\in[-\infty,\infty]$.

That the negativity of this function is infinite is demonstrated by
starting from the observation that the integral of negativity over $p$ 
from $-\infty$ to $\infty$ will diverge if the integral from any $p_0$ to 
$\infty$ diverges. To simplify calculations in the following analysis of 
divergence, we will examine some $p_0>>\hbar n$ so that the $nx$ terms in 
equation (\ref{squarewig}) become negligible. The integral computing the 
negativity between $p_0$ and $\infty$ then becomes:
\begin{align}
\label{squareneg}
\frac{1}{\pi^{2}}\int_{p_0}^\infty\bigg(\int_0^{\pi/2}
\mathbf{Neg}[(1-\cos[2nx])\frac{\sin[2(p/\hbar)x]}{p}]dx+\notag
\\
\int_{\pi/2}^{\pi}\mathbf{Neg}[(1-\cos[2n(\pi-x)])
\frac{\sin[2(p/\hbar)(\pi-x)]}{p}]dx\bigg)dp
\end{align}
where
\begin{equation*}
\mathbf{Neg}[x]= 
\begin{cases} -x & \text{if $x<0$,}
\\
0 &\text{if $x\ge 0$.}
\end{cases}
\end{equation*}
A variable substitution in the second line of equation (\ref{squareneg}) 
further reduces it to:
\begin{equation}
\label{reducedsquare}
\frac{2}{\pi^2}\int_{p_0}^\infty\int_0^{\pi/2}\mathbf{Neg}[(1-\cos[2nx])
\frac{\sin[2px/\hbar]}{p}]dxdp.
\end{equation}
Since $(1-\cos[2nx])$ and $\frac{1}{p}$ will both always be positive over
the domain of integration, they can be pulled out of the {\bf Neg} operator, 
leaving the equation in the form
\begin{equation}
\frac{2}{\pi^2}\int_{p_0}^\infty\int_0^{\pi/2}(1-\cos[2nx])
\frac{{\bf Neg}[\sin[2px/\hbar]]}{p}dxdp.
\end{equation}
The average value of {\bf Neg}$[\sin[2px/\hbar]]$ over a single period
of oscillation is $\frac{1}{\pi}$.  As $p\rightarrow \infty$, the period of 
oscillation $\rightarrow$ 0, so {\bf Neg}$[\sin[2px/\hbar]]$ effectively 
becomes a constant multiplier of $\frac{1}{\pi}$ in the large-$p$ limit, 
yielding the integral
\begin{equation}
\label{itdiverges}
\frac{2}{\pi^3}\int_{p_0}^\infty\int_0^{\pi/2}(1-\cos[2nx])\frac{1}{p}dxdp
=\frac{1}{\pi^2}\int_{p_0}^\infty\frac{dp}{p}
\end{equation}
which is divergent.

It is certainly interesting that the negativity of this Wigner function is 
divergent, especially since the Wigner function itself is defined to be 
normalized. The fact that the infinite square well's negativity diverges 
while the harmonic oscillator's does not points at a fundamental difference 
between the two systems, and merits further consideration. Nevertheless, 
the divergence of equation (\ref{itdiverges}) makes unsmoothed negativity 
an unhelpful metric when dealing with square wells or any states in a 
finite position-space domain, since those can always be expressed as a 
superposition of square well states.

Coarse-graining the Wigner function before measuring negativity provides a 
solution to this difficulty. We can demonstrate this with the following
argument: Since the Wigner function (smoothed or not) is by definition 
finite and continuous, we need not worry about divergence of 
the negativity integral except over infinite regions. Moreover, since 
equation (\ref{squarewig}) is symmetric in $p$, the integral of negativity 
over all phase space will converge if the integral from $0\text{ to }\infty$ 
converges. Combining these two observations, we can see that when proving the 
convergence of negativity over the entire Wigner function, it is sufficient 
to show convergence from any $p_0\text{ to }\infty$. Here as before, we 
examine some $p_0>>\hbar n$ to simplify calculations. By applying the same 
steps to the smoothed Wigner function that we applied to the unsmoothed 
Wigner function in reaching equation (\ref{reducedsquare}), we obtain the 
expression:
\begin{equation}
\frac{2}{\pi^2}\int_{p_0}^\infty\int_0^\pi\mathbf{Neg}[\int_{p_0}^\infty
\int_0^\pi{}e^{-\delta((x-x')^2+(p-p')^2)}(1-\cos[2nx'])
\frac{\sin[2p'x'/\hbar]}{p'}dx'dp']dxdp
\end{equation}
which can be rewritten as
\begin{align}
\label{separated}
\frac{2}{\pi^2}\int_{p_0}^\infty\int_0^\pi\mathbf{Neg}
[\int_{p_0}^\infty\frac{1}{p'}e^{-\delta(p-p')^2}F(p',x)dp']dxdp
\\
\text{where }F(p',x)=\int_0^\pi{}e^{-\delta(x-x')^2}(1-\cos[2nx'])
\sin[2p'x'/\hbar]dx'
\notag
\end{align}
Based on numerical analysis with \emph{Mathematica}, $F(p',x)$ is an 
oscillating function of $p'$ with an amplitude that decreases as 
$\frac{1}{p'}$ for sufficiently large $p'$. Since the negative value of this 
function will always be $\leq\frac{k}{p'}$, where $k$ is some constant 
dependent on $x$ and $\delta$, Eq.~(\ref{separated}) must have a value 
less than or equal to:
\begin{equation}\label{soclose}
\frac{2}{\pi^2}\int_{p_0}^\infty\int_0^\pi\mathbf{Neg}[\int_{p_0}^\infty 
-\frac{k}{p'^2}e^{-\delta(p-p')^2}dp']dxdp
\end{equation}
Furthermore, after convolution with a Gaussian in $p$, the function 
$\frac{k}{p^2}$ still drops off as some constant over $p^2$, so 
Eq.~(\ref{soclose}) further reduces to:
\begin{equation}
\frac{1}{\pi^2}\int_{p_0}^\infty\int_0^\pi\frac{k}{p^2}dxdp
\end{equation}
which clearly converges. Since the value of this integral is greater than or 
equal to the integral of negativity from $p_0\text{ to }\infty$, this 
integral must converge. Thus, the integral of negativity over all phase 
space must converge. Note that this proof is independent of $\delta$; any 
degree of coarse-graining will allow the integral to converge.

Some of the coarse-grained negativities of an infinite square well of
length $\pi$ are shown in Fig. 4. For any $\delta$, negativity generally
increases with increasing $n$ over the range shown. However, this growth
is not monotonic; at $n=8$, negativity decreases slightly from $n=7$. It
is also worth noting that the negativities for $n=5\text{ and }n=6$ seem
to show a strong dependence on $\delta$ that is absent from the rest of
the graph. The significance of these features is at present unclear.
They may correspond with actual variations in a square well's properties
at different energy levels, or they may simply indicate that small
fluctuations in $\eta$ should be ignored when considering larger trends
in a system's Wigner function. In either case, comparisons of calculated
negativities with experimental square well behavior will be needed to
determine how much of a fluctuation in $\eta$ constitutes a significant
physical change. Note also that the coarse-grained Wigner function
becomes positive definite for $\delta =1$, which, in the units used for
this problem, corresponds to the minimum uncertainty Gaussian again.
\begin{figure}[htp]
\centering
\includegraphics{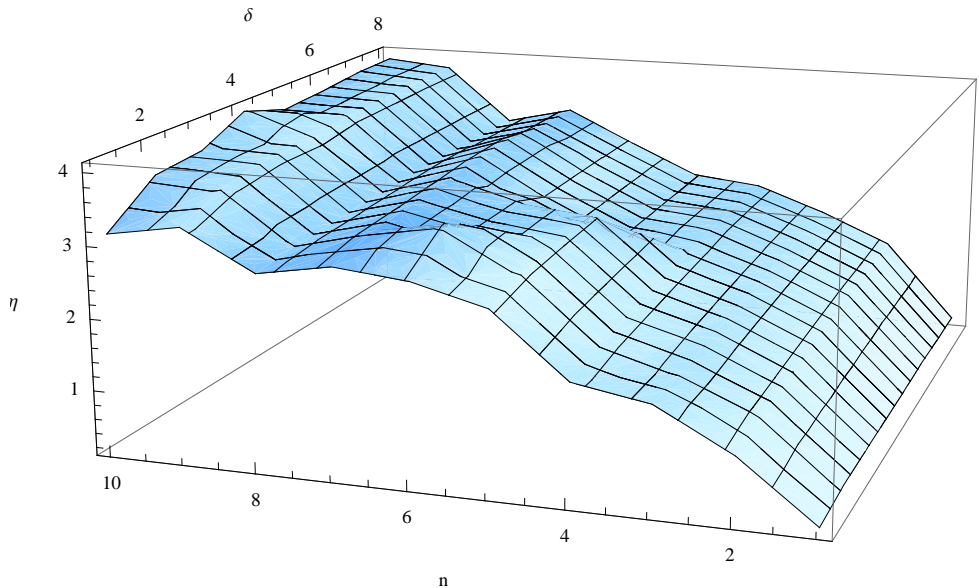}
\caption{The negativity $\eta$ of an infinite square well's smoothed Wigner 
function at various $n$ and $\delta$. Negativity seems to increase,
albeit non-monotonically, with $n$ for all $\delta$.}
\label{well}
\end{figure}

\section{Discussion}
While it is satisfying to see that coarse-graining the Wigner function
before computing its negativity resolves some critical issue with this
measure of quantumness, it invites the question of the physical 
interpretation of $\delta$. That is, what determines the degree of 
theoretical coarse-graining applied to a given Wigner function? One 
obvious consideration is that $\delta$ should be understood as a measure 
of the amount of thermal noise present that is affecting the system, whence 
the coarse-graining represents the `washing out' of small quantum features 
by thermal fluctuations. Under this interpretation, the increase 
in $n_{max}$ with increasing $\delta$ would correspond to the fact that 
larger systems can exhibit quantum behavior at sufficiently low 
temperatures, as is intuitive, while still providing for an appropriate
'correspondence principle' behavior for sufficiently high $n$. 

Alternatively, $\delta$ could be based on the precision of whatever 
measurements are being taken on the system. In this case, $\eta$ would be 
less an index of the quantumness \emph{present} in a system and more an 
index of how much of a system's quantumness could be observed given a 
certain precision of measurement. This interpretation could be useful in 
examining the classical-to-quantum transition as one 'zooms out' from the 
Planck scale and sacrifices small-scale precision for a more macroscopic 
view. It could also be useful in searching for macroscopic quantum behavior: 
Systems showing high negativity even after strong coarse-graining might 
be expected to show quantum behavior even on large scales.

\end{document}